\documentclass[%
 reprint,
 showpacs,
 noeprint,
 amsmath,amssymb,
 aps,
 prb,
 citeautoscript,
]{revtex4-1}
\usepackage[dvipdfmx]{graphicx}
\usepackage{dcolumn}
\usepackage{bm}

\begin{document}


\title{Thermodynamic properties of the $S=1/2$ twisted triangular spin tube}

\author{Takuya Ito}
 \email{ito@cmpt.phys.tohoku.ac.jp}
\author{Chihiro Iino}
\author{Naokazu Shibata}%
\affiliation{%
 Department of Physics, Tohoku University, Sendai 980-8578, Japan
 }%


\date{\today}

\begin{abstract}
Thermodynamic properties of the twisted three-leg spin tube under magnetic field are studied by the finite-$T$ density-matrix renormalization group method. The specific heat, spin, and chiral susceptibilities of the infinite system are calculated for both the original and its low-energy effective models. The  obtained results show that the presence of the chirality is observed as a clear peak in the specific heat at low temperature and the contribution of the chirality dominates the low-temperature part of the specific heat as the exchange coupling along the spin tube decreases. The peak structures in the specific heat, spin, and chiral susceptibilities are strongly modified near the quantum phase transition where the critical behaviors of the spin and chirality correlations change. These results confirm that the chirality plays a major role in characteristic low-energy behaviors of the frustrated spin systems.
\end{abstract}
\maketitle


\section{Introduction}

Geometrical frustration as a source of diverse and complex behaviors of electrons is an interesting research subject of condensed matter physics\cite{Kawamura1998}.
In quantum spin systems the chiral degrees of freedom at low temperatures play an important role in complicated behaviors\cite{Okunishi2008}. The antiferromagnetic Heisenberg models composed of weakly coupled triangular units are typical systems in which three $S=1/2$ spins in each triangular unit form a fourfold ground-state multiplet consisting of spin and chiral degrees of freedom\cite{Sakai2010}. The appearance of chiral degrees of freedom at low temperature is reflected in the characteristic behaviors of thermodynamic quantities and its critical correlation finally causes various quantum phase transitions at zero temperature.

\begin{figure}[t]
\centering
\includegraphics[width=8.6cm]{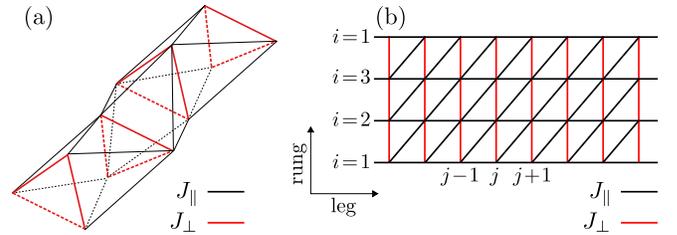}
\caption{(a) Twisted three-leg spin tube. The red lines represent $J_\perp$ that form unit triangles composed of $S=1/2$ quantum spins. They are connected by $J_\parallel$ of the black lines. (b) Unfolded diagram of the twisted three-leg spin tube. The lattice structure is equivalent to the triangular lattice.}
\label{figure1}
\end{figure}

\begin{figure}[t]
\centering
\includegraphics[width=8.6cm]{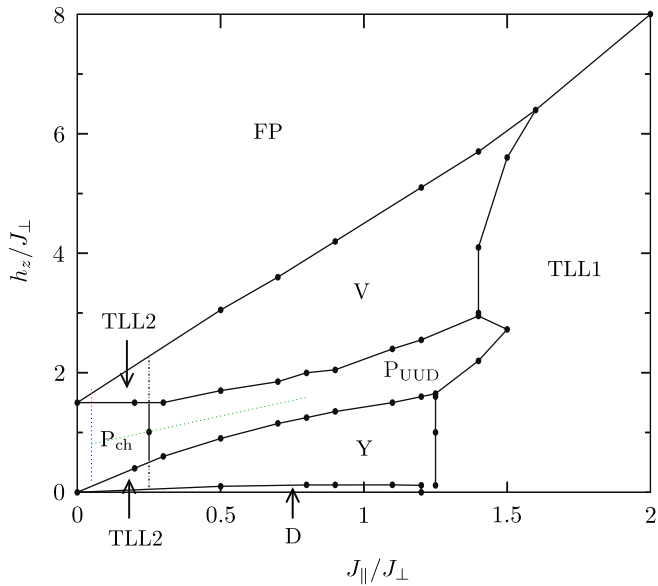}
\caption{Ground-state phase diagram of the twisted three-leg spin tube\cite{Yonaga2015}: D, dimer state; TLL2, two-component Tomonaga-Luttinger (TL) liquid; P$_{\rm ch}$, chirality liquid in the magnetization plateau; Y, Y phase of the triangular lattice; UUD, up-up-down (UUD) phase; V, V phase of the triangular lattice; TLL1, one-component TL liquid; and FP, fully polarized state. The blue, red and green dashed lines indicate parameters used in Figs. 4(a), 5(a) and 7(a), respectively.}
\label{figure2}
\end{figure}

The twisted three-leg spin tube shown in Fig.~\ref{figure1} is one of the simplest one-dimensional systems whose low-energy states are described by both spin and chirality of the triangular unit\cite{Schnack2004,Ivanov2010,Furukawa2011}. The degeneracy in each triangular unit is lifted by the exchange coupling along the spin tube and many quantum states including chirality liquid, spin liquid, dimer state, and spin-gap state are realized depending on the exchange coupling and the magnetic field as shown in Fig.~\ref{figure2}\cite{Cabra1998,Tandon1999,Orignac2000,Luscher2004,
Yoshikawa2004,Okunishi2005,Fouet2006,Sakai2010,Chen2013,Yonaga2015}.

In this paper we study thermal properties of the twisted three-leg spin tube whose low-energy states are described by spin and chiral degrees of freedom.
We calculate the partition function of the system by using the finite-$T$ density-matrix renormalization group (DMRG) method \cite{White1992,White1993,Wang1997,Shibata1997,Sirker2002a} and directly determine the specific heat, spin, and chiral susceptibilities using the eigenvectors of the maximum eigenvalue of the quantum transfer matrix\cite{Betsuyaku1984,Betsuyaku1985}. The obtained results for the infinite system show that the presence of the chirality is observed as a peak structure in the specific heat and the contribution from the chirality becomes dominant at low temperature as the exchange coupling along the spin tube decreases. The peak structures in the specific heat and the susceptibilities at low temperatures are strongly modified near the quantum phase transitions where the critical behaviors in the ground state change. 

\section{Model and method}
The Hamiltonian of the twisted three-leg spin tube is defined as
\begin{eqnarray}
H &=&  J_{\perp}{\displaystyle \sum_{j=1}}{\displaystyle \sum_{i=1}^3} \bm{S}_{i,j} \cdot \bm{S}_{i+1,j} - h_{z}{\displaystyle \sum_{j=1}}S^{z}_{{\rm T}j} \nonumber \\
&+& J_{\parallel}{\displaystyle \sum_{j=1}}{\displaystyle \sum_{i=1}^3}(\bm{S}_{i,j} \cdot \bm{S}_{i,j+1} + \bm{S}_{i,j} \cdot \bm{S}_{i+1,j+1}),
\label{eqHamiltonian}
\end{eqnarray}
where $\bm{S}_{i,j}$ is the $S\!=\!1/2$ spin operator at rung $i$ and leg $j$,
and $J_{\perp}\ (J_{\parallel})$ is the antiferromagnetic exchange coupling of
intra (inter) triangles. The second term represents the Zeeman energy under the external magnetic field $h_z$. The total spin of the $j$th unit triangle is defined by
$\bm{S}_{{\rm T}j} = \sum_{i=1}^3 \bm{S}_{i,j}$.
Here we use the periodic boundary conditions for
the perpendicular direction to the tube axis, and we regard $J_{\perp}$ as energy unit throughout this paper.

To calculate thermodynamic quantities of the Hamiltonian
we use the finite-$T$ DMRG method\cite{White1992,White1993,Wang1997,Shibata1997,Sirker2002a}.
This method enables us to iteratively expand the quantum transfer matrix\cite{Betsuyaku1984,Betsuyaku1985} in the imaginary time direction and obtain the partition function of the infinite system from the maximum eigenvalue of the quantum transfer matrix within a desired accuracy controlled by the number of keeping states, $m$, in the calculation.
We treat each unit triangle of the spin tube as a single site and keep up to 300 basis states ($m=300$) in each block of the quantum transfer matrix.
The accuracy of the results depends on the number of keeping states $m$ and the Trotter number $N_{\rm T}$ of the quantum transfer matrix used in the calculation\cite{Wang1997,Shibata1997}. In the present calculation the error coming from finite $m$ is in a range of $O(10^{-5})$ to $O(10^{-12})$ and the error from the finite Trotter number is removed by using an extrapolation of $N_{\rm T}$ in a range of $N_{\rm T}=50$ -- $200$. Hereafter we set $k_B$ as 1.

\begin{figure}[t]
\centering
\includegraphics[width=8.6cm]{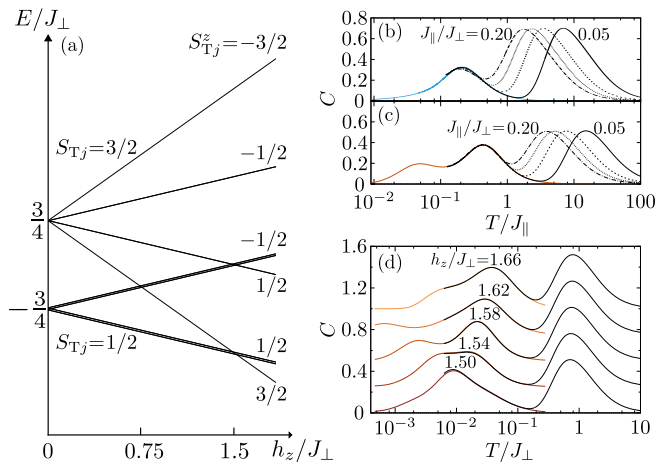}
\caption{(a) Energy levels of the triangular unit composed of three $S=1/2$ spins connected by the antiferromagnetic coupling $J_\perp$ under the magnetic field $h_z$. Double lines mean the presence of twofold degeneration of chirality. (b), (c) Specific heat of the original twisted triangular spin tube per unit triangle (solid and dashed black lines) for $J_{\parallel}/J_{\perp}=0.05, 0.10, 0.15, 0.20$ from right to left and that of the effective model (color) defined by Eqs.~(\ref{eqHplateau}) and (\ref{eqHtJ}). The applied magnetic field is chosen as $h_z/J_{\perp} = 0.75 + J_{\parallel}/J_{\perp}$ and $h_z/J_{\perp} = 1.5 + 1.6\times J_{\parallel}/J_{\perp}$, respectively. (d) Specific heat of the original twisted triangular spin tube per unit triangle (black) and that of the effective model (color) defined by Eq.~(\ref{eqHtJ}) under various magnetic field at $J_{\parallel}/J_{\perp}=0.05$.
}
\label{figure3}
\end{figure}

\section{Effective Hamiltonian}
We start from the decoupling limit $J_{\parallel} = 0$, where the unit triangles perpendicular to the tube axis are independent of each other.
The eigenstates in this limit are obtained by diagonalizing an $8\times 8$ Hamiltonian of three $S=1/2$ spins in the unit triangle and they are shown to form two fourfold multiplets in the absence of a magnetic field as in Fig.~\ref{figure3}(a). The higher multiplet has total spin $S_{{\rm T}j}=3/2$ and the lower one is composed of $S_{{\rm T}j}=1/2$ states with chiral degrees of freedom, which are represented by pseudospin $\tau^z=\pm 1/2$. The degeneracy in each multiplet is partially lifted by Zeeman energy and only chiral degrees of freedom remain in the ground state under magnetic field until a large Zeeman energy fully polarizes the spins.

The exchange interaction along the spin tube $J_\parallel$ generates matrix elements between the eigenstates of unit triangles, and the effective Hamiltonian acting on the chiral degrees of freedom is derived within the first-order perturbation with respect to $J_\parallel$ as
\begin{equation}
H_{\rm{XY}} =-\frac{J_{\parallel}}{3} \sum _{j=1} (\tau ^{+}_{j}\tau ^{-}_{j+1} + {\rm H.c.}),
\label{eqHplateau}
\end{equation}
where $\tau^+_j(\tau^-_j)$ is the chirality pseudospin-1/2 raising (lowering) operator\cite{Fouet2006,Schulz1996,Kawano1997}.
In the presence of a finite spin excitation gap in the magnetization plateau phase, the spin excitations are suppressed and a chirality liquid state is realized at low temperature. Thus the low-energy thermodynamic properties in the magnetization plateau phase are expected to be characterized by the chirality effective Hamiltonian $H_{\rm{XY}}$. This is confirmed by comparing the specific heat of both the original and effective Hamiltonians shown in Fig.~\ref{figure3}(b), where we find almost identical results below $T/J_{\parallel} \sim 1$, which means the effective Hamiltonian of chiral degrees of freedom reproduces the original specific heat at low temperatures.

With the increase of the magnetic field, the energy gap of the spin excitation decreases and finally vanishes at the transition to the spin and chirality liquid state. This change in the ground state is due to the large Zeeman shift of the $S_{{\rm T}j}^z=3/2$ state that modifies the effective Hamiltonian for the chiral degrees of freedom as
\begin{eqnarray}
{H}_{{\rm t \mathchar`-J}} = {\displaystyle \sum _{j=1}}{\displaystyle \sum _{\sigma = \uparrow ,\downarrow}} \Bigl[ -t(c^{\dagger}_{\sigma j}c_{\sigma j+1} &+& {\rm H.c.}) 
- J(\tau ^{+}_{j}\tau ^{-}_{j+1} + {\rm H.c.}) \nonumber \\ &+& Vn_{j}n_{j+1} + \mu n_{j} \Bigr],
\label{eqHtJ}
\end{eqnarray}
where $c^{\dagger}_{\sigma j}$ ($c_{\sigma j}$) and $n_j$ are hard-core boson creation (annihilation) operator with chirality pseudospin-1/2 and its number operator, respectively\cite{Tandon1999,Citro2000}. The coefficient of each term is given as $t=\frac{1}{2}J_{\parallel}$, $J=\frac{1}{3}J_{\parallel}$, $V=\frac{2}{3}J_{\parallel}$, and $\mu = -2J_{\parallel}+(h_{z} - \frac{3}{2}J_{\perp})$.
This effective Hamiltonian corresponds to the extended $t$-$J$ model with XY exchange coupling. The field-induced phase transition from the plateau phase to the spin and chirality liquid phase then corresponds to the metal-insulator transition in the $t$-$J$ model\cite{Ogata1991}.
The equivalence between the effective $t$-$J$ model and the original spin tube in the low-energy region is shown in Fig.~\ref{figure3}(d), where we find that the specific heat of the original spin tube is well reproduced by the effective Hamiltonian at low temperatures. This agreement between the two models is seen even for $J_\parallel/J_\perp=0.2$ as shown in Fig.~\ref{figure3}(c).

\section{Results}
\subsection{Chirality liquid state}
\begin{figure}[t]
  \centering
   \includegraphics[width=8.6cm]{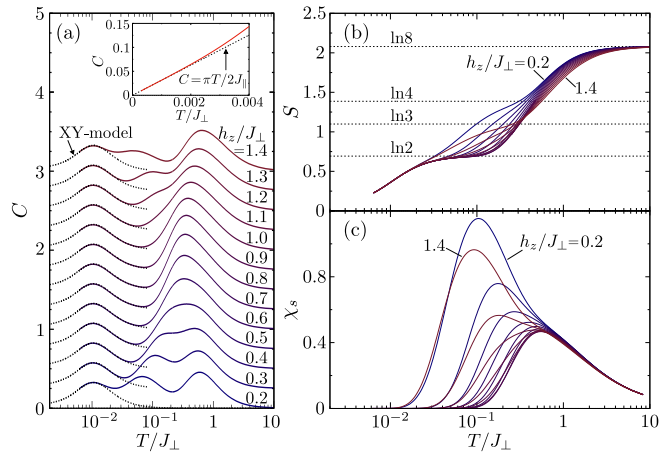}
   \caption{(a) Specific heat of twisted triangular spin tube per unit triangle at $J_\parallel/J_\perp = 0.05$ and that of effective XY model defined in Eq.~(\ref{eqHplateau}) under various magnetic field. (b) Entropy per unit triangle and (c) spin susceptibility of the twisted triangular spin tube. The same line color in the figures represents the same magnetic field. Inset in (a) shows asymptotic behavior of $C_{\rm XY}$ at low temperatures obtained by the present calculation (red line), which is consistent with the prediction of CFT ($C=\pi T/2J_{\parallel}$) \cite{Blote1986,Affleck1986} represented by the broken line.}
	\label{figure4}
\end	{figure}

To clarify the thermodynamic behaviors of chirality as effective degrees of freedom of the unit triangle, we first investigate the specific heat $C$ of the twisted three-leg spin tube in the region of the plateau phase (P$_{\rm ch}$ in Fig.~\ref{figure2}).
The results for $J_\parallel/J_\perp=0.05$ under various magnetic field are presented in Fig.~\ref{figure4}(a), where we find at least two peaks, one of whose position at lower temperature is independent of the magnetic field. The presence of the magnetic-field-independent peak at low temperature is consistent with the expectation that the chirality has the smallest energy scale given by the effective Hamiltonian of Eq.~(\ref{eqHplateau}). The other peaks at higher temperature are magnetic field dependent and their positions correspond to the excitation energies to excited states of different $S_{{\rm T}j}^z$ and $S_{{\rm T}j}$ shown in Fig.~\ref{figure3}(a).
The specific heat $C_{\rm XY}$ of the effective Hamiltonian consisting of only chirality operators in Eq.~(\ref{eqHplateau}) is represented by the dashed line in the same figure whose low-temperature asymptotic behavior is consistent with the prediction of conformal field theory (CFT) ($C=\pi T/2J_{\parallel}$),\cite{Blote1986,Affleck1986} as 
shown in the inset of Fig.4 (a). The almost identical results between $C$ and $C_{\rm XY}$
at low temperatures show that the low-energy properties are described by chiral degrees of freedom. This coincidence of the specific heat also indicates that the energy scale of chirality is well separated from that of spin excitations as is shown in the temperature dependence of the entropy in Fig.~\ref{figure4}(b), where the plateau of $S\sim \ln2$ coming from the chiral degree of freedom appears in the intermediate temperature range between the lowest and the second lowest peaks of the specific heat around $h_z/J_\perp=0.8$, while different plateaus of $\ln 4$ and $\ln 3$ appear near $h_z/J_{\perp}=0.2$ and $h_z/J_\perp=1.4$ that correspond to the lower and upper boundary of the plateau phase, where different local spin states with $S_{{\rm T}j}^z=-1/2$ and $3/2$ are included in the ground state, respectively. The presence of the large spin excitation gap around $h_z/J_\perp=0.8$ is also confirmed in the temperature dependence of the spin susceptibility shown in Fig.~\ref{figure4}(c), where we find vanishing spin susceptibility below the intermediate-temperature region $T/J_\perp \sim 0.1$. From these results we conclude that the presence of the chirality in the spin tube is observed in the specific heat at low temperature.

\begin{figure}[t]
\centering
\includegraphics[width=8.6cm]{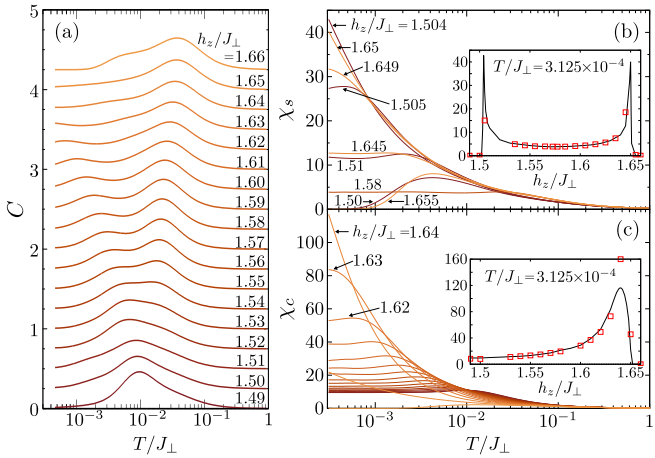}
\caption{(a) Specific heat per unit triangle, (b) spin susceptibility, and (c) chiral susceptibility of the effective Hamiltonian of the twisted triangular spin tube [Eq.~(\ref{eqHtJ})] under various magnetic field at $J_{\parallel}/J_{\perp}=0.05$. The chiral susceptibility is defined by $\chi_c = \lim_{\delta \rightarrow 0} d\langle \tau_j^z \rangle/d\delta$ under small chirality Zeeman energy $\delta\sum_j \tau^z_j$. Insets in (b) and (c) show field dependence of the susceptibilities  at $T/J_{\perp}=3.125 \times 10^{-4}$. Open squares represent the results obtained by DMRG calculations with sine square defromeaiton\cite{Gendiar2009,Gendiar2009a,Hotta2012} at $T/J_{\perp}= 0$. The same line color in the figures represents the same magnetic field.}
\label{figure5}
\end{figure}

\begin{figure}[t]
\centering
\includegraphics[width=8.6cm]{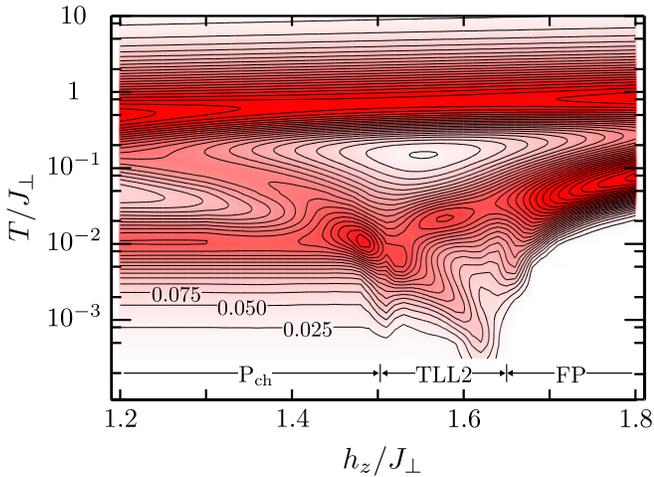}
\caption{Contour map of the specific heat per unit triangle around the spin and chirality liquid (TLL2) phase at $J_{\parallel}/J_{\perp}=0.05$. The high-temperature region is obtained in the original model. The interval of contour lines is $0.025$. Unusual temperature dependence is seen at the critical magnetic fields $h_z/J_{\perp}=1.5$ and $1.65$.}
\label{figure6}
\end{figure}

\begin{figure}[t]
\centering
\includegraphics[width=8.6cm]{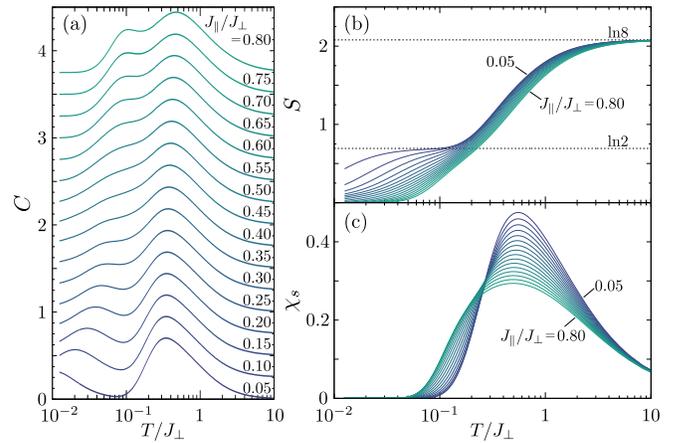}
\caption{$J_{\parallel}/J_{\perp}$ dependence of (a) specific heat per unit triangle, (b) entropy per unit triangle, and (c) spin susceptibility in the magnetization plateau. To go through around the center of the plateau, we applied magnetic field as $h_z/J_{\perp} = 0.75 + J_{\parallel}/J_{\perp}$. The same line color in the figures represents the same exchange couplings.}
\label{figure7}
\end{figure}

\subsection{Spin and chirality liquid state}
We next examine the interplay between the spin and chirality in the TLL2 phase at higher magnetic field, where both spin and chiral excitations are described by gapless collective excitations. Since the Zeeman energy enhances the probability of the spin-polarized $S_{{\rm T}j}^z=3/2$ states (holes in the effective $t$-$J$ model), its effect on the chiral excitations is expected. Here we study the effective Hamiltonian Eq.~(\ref{eqHtJ}) of the spin tube to see how the specific heat of chirality is modified with the increase of the magnetic field.

At $h_z/J_{\perp}\sim 1.49$, the ground state is in the plateau phase (${\rm P_{ch}}$) and we see a single peak at low temperature in Fig.~\ref{figure5}(a), which continuously connects to the specific heat of the chirality XY model, Eq.~(\ref{eqHplateau}), at lower magnetic field shown in Fig.~\ref{figure4}(a). After the phase transition at $h_z/J_{\perp}=1.5$, the number of spin-polarized $S_{{\rm T}j}=3/2$ states starts to increase with the magnetic field. This increase of the spin-polarized states enhances a gapless spin excitation at finite temperature and a small increase of the specific heat appears at low temperatures $T/J_{\perp} \sim 0.002$. Further increase in magnetic field modifies the peak structure of chiral excitations and it finally falls into decay toward lower temperature. The peak structure of chiral excitations finally seems to vanish at $h_z/J_{\perp}=1.65$, which corresponds to the transition field to the fully spin-polarized state. This field dependence of the specific heat is clearly seen in Fig.~\ref{figure6}, which shows a contour map of $C$ around the TLL2 phase.

To understand the magnetic field dependence of the specific heat,
we next investigate the spin and chiral susceptibilities.
Figures \ref{figure5}(b) and \ref{figure5}(c) show the temperature dependence of the spin and chiral susceptibilities, $\chi_s$ and $\chi_c$. We find that the low-temperature chiral susceptibility monotonically increases with the increase of magnetic field up to the transition field of $h_z/J_{\perp}=1.65$. This enhancement of the chiral susceptibility is due to the increase in the number of spin-polarized $S_{{\rm T}j}=3/2$ states because the effective interaction between the chirality operators monotonically decreases with the increase of the number of spin-polarized $S_{{\rm T}j}=3/2$ states and this decrease of the interaction results in the enhancement of the chiral susceptibility.
A similar sharp increase appears in the spin susceptibility at the transition to the TLL2 phase at $h_z/J_{\perp}=1.5$ at low temperature. This increase of the spin susceptibility originates from the appearance of low-energy spin excitations of the spin-polarized $S_{{\rm T}j}=3/2$ states. The increase in the number of spin-polarized $S_{{\rm T}j}=3/2$ states first enhances the possibility of the exchange process and the energy of the spin excitations increases, which suppresses the spin susceptibility at low temperature. However, further increase of the magnetic field suppresses the exchange processes between $S_{{\rm T}j}^z=3/2$ and $1/2$ states and the energy scale of the spin excitations decreases, which enhances the spin susceptibility near the phase transition to the fully spin-polarized states.
Thus the behaviors of the spin and chirality are significantly modified by the presence of the spin-polarized $S_{{\rm T}j}=3/2$ state, whose number is controlled by the magnetic field.

\subsection{Up-up-down state}

We finally study how the exchange coupling between the neighboring triangular units modifies the thermodynamic behaviors of the spin tube.
We start from the chirality liquid state in the magnetization plateau phase to see the effect of interactions between the chiral degrees of freedom. The $J_{\parallel}$ dependence of the specific heat of the triangular spin tube is shown in Fig.~\ref{figure7}(a). In the weak-coupling region of $J_{\parallel}/J_{\perp} < 0.25$, we find a clear peak at low temperature where magnetic susceptibility is suppressed. The released entropy associated with this peak formation is close to $\ln 2$ per unit triangle as shown in Fig.~\ref{figure7}(b), which shows this peak in $C$ is coming from chiral degrees of freedom. With increasing $J_{\parallel}/J_\perp$, the peak structure in $C$ shifts to higher temperature and merges with the main peak at higher temperature, whose origin is the spin degrees of freedom of the original spins in triangular units. The lower peak is deformed into a shoulder structure at around $J_\parallel/J_\perp=0.25$ and then
a small peak is formed again. This peak shift and transient behavior of the specific heat implies a reconstruction of the excitation spectrum at $J_\parallel/J_\perp\sim 0.25$, which corresponds to the phase transition from the ${\rm P_{ch}}$ phase to the up-up-down (UUD) phase in Fig.~\ref{figure2}.
In particular the vanishing specific heat at low temperatures means the disappearance of low-energy chiral excitations characterizing the chirality liquid phase. Because the spin excitation gap is always present in the magnetization plateau phase as is seen in a vanishing low-temperature spin susceptibility in Fig.~\ref{figure7}(c), the low-energy thermodynamic properties are scaled by finite energy gaps of both the spin and chiral excitations. This change in the low-energy chiral excitation is consistent with the presence of the phase transition from chirality liquid to UUD phase in the ground state\cite{Yonaga2015}.

\section{Summary}

In this paper we have studied thermal properties of the twisted three-leg spin tube by using the finite-$T$ DMRG method. The obtained specific heat in the magnetization plateau state shows a clear peak structure independent of the magnetic field, indicating the presence of the chiral degrees of freedom at low temperature. With the increase in magnetic field, spin-polarized local states with no chirality are included in the ground state with the phase transition to the TLL2 phase. The spin susceptibility sharply increases at the transition at low temperature and then decreases with the increase in the number of spin polarized local states up to around half the number of unit triangles. The chiral susceptibility gradually increases with the increase in the number of spin-polarized states and shows diverging behavior in the limit of low temperature at the transition to the fully spin-polarized state.
We have also studied the effect of interactions $J_{\parallel}$ between the triangular unit in the spin tube. With the increase of the exchange interaction between the spins in neighboring triangular units, the temperature dependence of the specific heat shows that the energy scale of the chiral excitations increases and the excitation gap finally appears with vanishing chiral susceptibility at low temperatures, consistent with the transition to the UUD ordered phase in the ground state. These results confirm that the chirality is an important physical quantity characterizing low-energy thermal properties of the twisted spin tube as a highly frustrated quantum spin system.

\begin{acknowledgments}
This work was supported by JSPS (JP) Grant No. 26400344.
\end{acknowledgments}

\bibliographystyle{apsrev4-1}
\bibliography{bibliography}

\end{document}